\documentclass{vldb}
\usepackage{graphicx}
\usepackage{balance}  
\usepackage{enumitem}
\usepackage{subfigure}
\usepackage[hyphens]{url}
\usepackage{hyperref}
\usepackage{xspace} 

\usepackage{color}

\newcommand\Mark[1]{\textsuperscript#1}
\newcommand{\systemname}{Spitz\xspace}

\vldbTitle{\systemname: A Verifiable Database System}
\vldbAuthors{Meihui Zhang, Zhongle Xie, Cong Yue, Ziyue Zhong}
\vldbDOI{https://doi.org/10.14778/3415478.3415567}
\vldbVolume{13}
\vldbNumber{12}
\vldbYear{2020}

\begin{document}


\title{\systemname: A Verifiable Database System}



%
%
%
%

\numberofauthors{1} 
\author{
%
%
\alignauthor 
Meihui Zhang\Mark{1} \qquad Zhongle Xie\Mark{2} \qquad Cong Yue\Mark{2} \qquad Ziyue Zhong\Mark{1} \\
\vspace{2mm}
\affaddr{\Mark{1} Beijing Institute of Technology \qquad \Mark{2} National University of Singapore}
\email{\tt meihui\_zhang@bit.edu.cn, zhongle@comp.nus.edu.sg, yuecong@comp.nus.edu.sg, ziyue\_zhong@bit.edu.cn}
}

\maketitle

\begin{abstract}

Databases in the past have helped businesses maintain and extract insights from their data. Today, it is common for a business to involve multiple independent, distrustful parties. This trend towards decentralization introduces a new and important requirement to databases: the integrity of the data, the history, and the execution must be protected. In other words, there is a need for a new class of database systems whose integrity can be verified (or verifiable databases).  


In this paper, we identify the requirements and the design challenges of verifiable databases.
We observe that the main challenges come from the need to balance data immutability, tamper evidence, and performance. We first consider approaches that extend existing OLTP and OLAP systems with support for verification. We next examine a clean-slate approach, by describing a new system, \systemname, specifically designed for efficiently supporting immutable and tamper-evident transaction management. We conduct a preliminary performance study of both approaches against a baseline system, and provide insights on their performance.

\end{abstract}

\section{Introduction}\label{sec:intro}

Traditional database systems are indispensable for businesses.They excel at storing, processing, and performing analytics over business transactions. Recent digital optimization and transformation have enabled businesses to transact directly with each other, without relying on a central party. As a result, multiple parties can access a shared database. Since the parties are mutually distrustful, the underlying database must consider support for auditing, tamper evidence, and dispute resolution in its design. For instance, it must maintain a trusted data history and allow users to verify the integrity of both current and historical data. 


Blockchains demonstrate one practical design for database systems with strong integrity~\cite{DBLP:journals/tkde/DinhLZCOW18}. Public blockchain systems, such as Ethereum, support secure peer-to-peer applications through smart contracts. Private blockchains, such as Hyperledger Fabric, target business settings and achieve higher performance than public ones.  
Blockchains have drawn interests from banks and regulators, with the prospect of offering digital currency and digital banking. Combined with recent advances in 5G, AI and IoT, block-chains are expected to speed up  the transformation and further disrupt the e-commerce and financial industries. 

To make a database that can be accessed by potentially malicious parties trustworthy, it must be {\em verifiable}. A verifiable database system protects integrity of the data, of its provenance, and of its query execution. More specifically, any tampering such as changing the data content, changing a historical record, or modifying query results, can be detected. 
We note that the demand for verification is on the rise due to the requirements imposed by the regulators on various business sectors, investment and banking in particular.


The first requirement in the design of a verifiable database (VDB) is data immutability, which is necessary for maintaining trusted provenance. Immutability means data is only written once and never deleted. It is not a new concept. It has been used in NoSQL systems such as HBase~\cite{hbase}, CouchDB~\cite{couchdb} and RethinkDB~\cite{rethinkdb} to achieve more efficient concurrency, due to the fact that no synchronization of accesses is needed. It is also used in Resilient Distributed Datasets (RDDs)\cite{spark} for lineage and fault tolerance. The second requirement of a verifiable database is query verifiability. It means the query results contain integrity proofs for both the data and query execution. More specifically, a user can detect if either the data or the query execution has been tampered. 



\begin{figure}[t]
  \centering
  \includegraphics[width=0.47\textwidth]{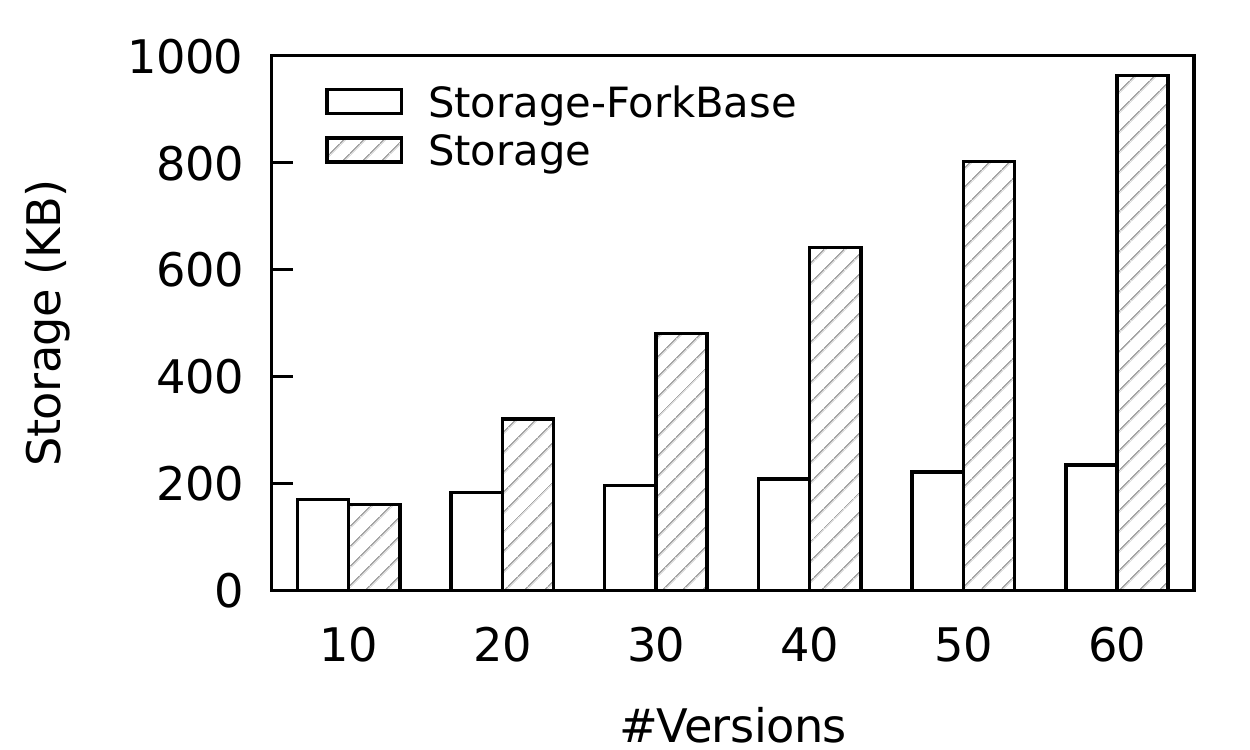}
  \caption{Data storage improved by deduplication.}
  \label{fig:dedup_time_space}
\end{figure}

In this paper, we discuss four challenges in realizing the design of verifiable databases. 
The first challenge is in storage management, as data immutability requires managing the ever-increasing volume of data. 
Consider a typical healthcare analytic application, in which health data needs to be kept for the lifetime of a patient, and each diagnosis, lab test, prescription, etc., is appended to the patient profile.  
Disease and procedure coding standards evolve over time, e.g., from ICD-9-CM to ICD-10 in recent years.  
Such changes in classification and coding standards require updates or mapping onto the existing medical record.  
To ensure good data provenance, the data must be immutable and a new version of the database, i.e., a snapshot, is appended.
The data volume is increasing with time, and therefore its management needs to be efficient and reliable.
Let us consider another example where an immutable database stores 10 WIKI pages of 16 KB each initially. We create a new version when updating a page, while keeping the previous versions.
Figure~\ref{fig:dedup_time_space} shows the space utilized with an increasing number of versions. Clearly, the space utilization increases substantially with the number of immutable versions, and the use of an efficient multi-version storage engine such as ForkBase~\cite{DBLP:journals/pvldb/WangDLXZCCOR18} helps to reduce it.
This highlights the importance of storage efficiency for a database that is forever increasing in size.

The second challenge is to provide efficient access methods for querying immutable data. While some existing database systems archive historical data and support temporal query processing, they have not been designed to support ``permanent" immutability. In VDB, data is never deleted; query processing and frequent searching on older versions of the data will be prohibitively expensive if efficient storage layout and indexes are not supported. This may entail scanning of a substantial portion of the database for answering verification queries.

The third challenge is to minimize performance overhead of verification. VDB must generate integrity proofs whose cost can be significant. Blockchains, for example, have poor transaction throughput due to their protocols for guaranteeing security in the Byzantine environment.  The performance gap between traditional databases and blockchains is significant due to their different design focus. As a result, a verifiable database must adopt a hybrid blockchain-database approach in order to strike a better balance between performance and security. 


The fourth challenge is the need to support both OLTP and OLAP workloads, as illustrated by the emergence of HTAP database systems. The former requires serializability which is important for applications such as e-commerce. Most existing OLTP systems adopt  optimistic concurrency control (OCC), instead of pessimistic concurrency control, because of its simplicity and high performance. In contrast, analytic queries in OLAP do not require the strict ordering provided by serializability. Existing OLAP systems adopt multi-version concurrency control (MVCC) to achieve data consistency with high performance. Most existing works on verifiable queries focus on OLTP workloads. While general OLAP queries can be made verifiable, for example by using fully-homomorphic encryption, they involve complex cryptographic operations and incur significant overhead \cite{ParnoHG013, DBLP:conf/uss/SettyVPBBW12}. Therefore, it is challenging to support both verifiable OLTP and OLAP queries with practical performance.

In addition to the four challenges above, we note that VDB must also aim for deployability. It is often costly to either add a new database system into an existing infrastructure, or to replace an existing database with a new one. 
In particular, for a business with consolidated software stacks, data conversion is necessary to move data to the new database.  Furthermore, users may find the system difficult to use if the verifiable database adopts unfamiliar programming models or interface.


\begin{figure}
    \centering
    \includegraphics[width=1\linewidth]{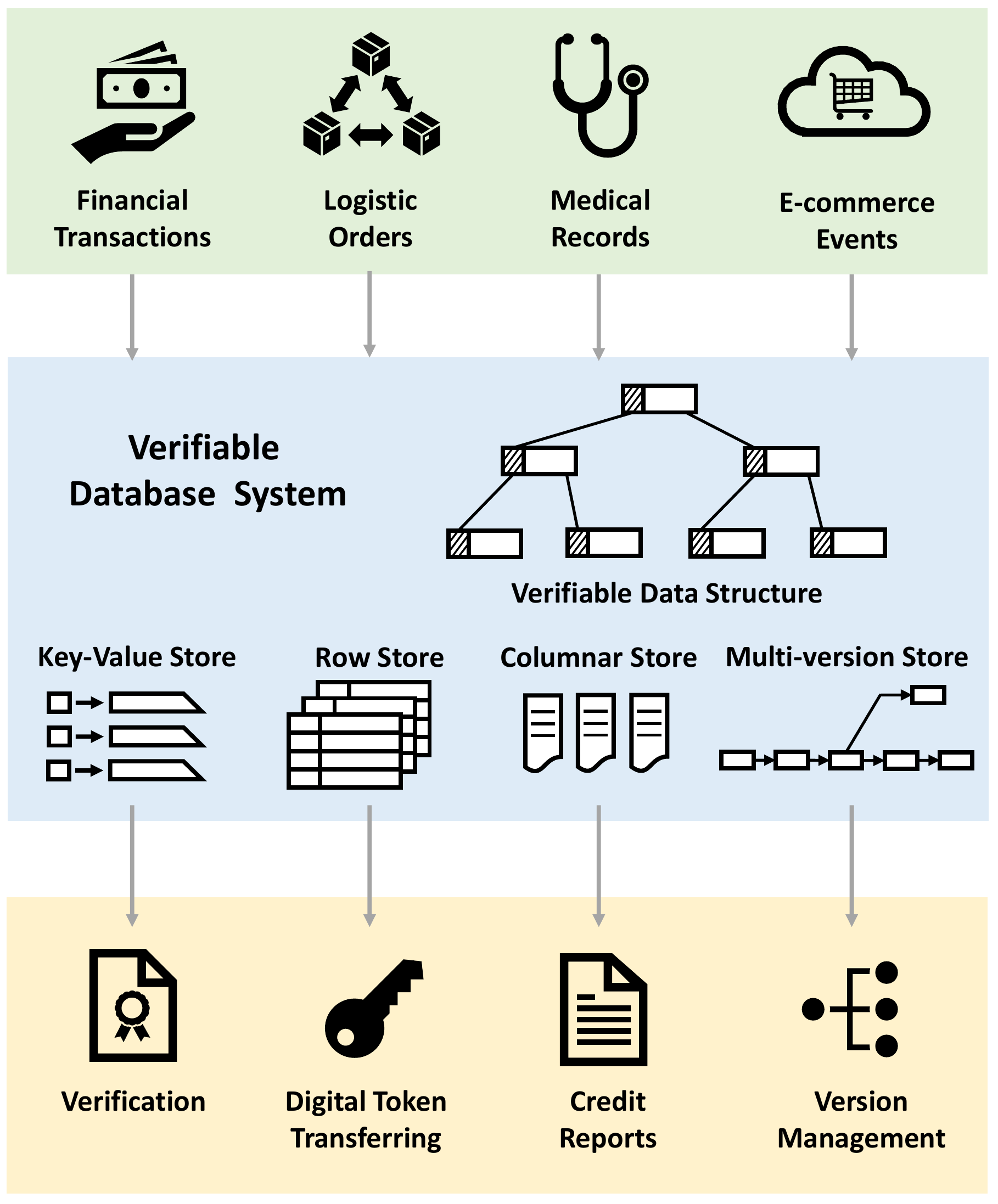}
    \caption{Verifiable database system overview.}
    \label{fig:system}
\end{figure}

In this paper, we discuss two approaches to realize an efficient VDB. The first approach is to extend existing systems, and the second is to design a new system from scratch. Figure ~\ref{fig:system} shows the second approach and how it fits with existing business applications. The new system uses tamper-evident structures for verification, and efficient version management for performance.  



We make the following contributions in this paper:
\begin{itemize}
    \item We identify the requirements and design challenges of efficient verifiable databases. 
    \item  We discuss two approaches for realizing an efficient verifiable database: by extending existing systems, and a new clean-slate design  called \systemname.  
    
    \item We perform an experimental study on \systemname and compare it with a baseline.
    The results show that \systemname can achieve good performance, despite overhead from verification  and additional data structures.
   \item
    We discuss various future research topics, including the integration of learning methods onto VDB and version management of the machine learning pipeline.
    
\end{itemize}

The rest of this paper is organized as follows.  In Section~\ref{sec:VDB}, we present existing works and systems that are related to VDB processing. In Section~\ref{sec:designissues}, we discuss the research challenges and opportunities of VDB. We next describe the challenges in the approach of extending existing OLTP and OLAP system to implement VDB in Section~\ref{sec:extension_approach}. In Section~\ref{sec:system},  we present the system architecture of \systemname. We then present an experimental study in Section~\ref{sec:exp}, and compare our systems against a baseline implementation based on a commercial service. We discuss the promising synergy between VDB and AI in Section~\ref{ai}, before concluding in Section~\ref{sec:conclude}.


\section{Verifiable Databases}
\label{sec:VDB}

In this section, we survey existing works and systems with verification features.

\subsection{Verifiable Database}

Data integrity is important in outsourced database as third-party service providers can be malicious. In particular, applications running on top of an outsourced database require the data and query results from the providers to be verifiable, that is, tampering of data and query execution can be securely detected. 
One way to achieve verifiability is using verifiable computation techniques.
Benabbas et al.~\cite{DBLP:conf/crypto/BenabbasGV11} present a delegation scheme on verifiable database minimizing the resources required by the clients of verifiable database.
Guo et al.~\cite{GuoLCW19} improve the update efficiency using a long polynomial for public keys and a short polynomial for private keys.
Miao et al.~\cite{MiaoWWM19} enable efficient keyword search for VDB using enhanced vector commitment while HVDB~\cite{ZhangCLTM19} supports hierarchical verification by building a vector commitment tree. 
SNARKs~\cite{ParnoHG013} can support arbitrary computation tasks, but requiring an expensive setup phase, and incurring significant overhead. Ben-Sasson et al.~\cite{BenSasson14} improve upon SNARKs by bounding the complexity of the setup phase to size of the database and the query complexity. 
More recently, Zhang et al.~\cite{ZhangGKPP17} propose a system called vSQL that uses an interactive protocol to support verifiable SQL queries. However, vSQL is limited to relational databases with a fixed schema.

Another way to achieve verifiability is by using authentication data structures such as Merkle trees.
Li et al.~\cite{LiHKR06} propose and evaluate authentication index structures combining Merkle trees and B$^+$-trees.
Yang et al.~\cite{YangPPK09} propose integrity-protected MR-tree for spatial data.
ServeDB ~\cite{servedb19} proposes a Merkle tree index based on hierarchical cube encoding that supports efficient multi-dimensional queries. 
Security conscious applications enforce data integrity against malicious modifications not only from external attackers, but also from malicious insiders and cloud hosting operators. 
As a solution,
SUNDR ~\cite{DBLP:conf/osdi/LiKMS04}
cryptographically protects all file system contents and proposes a fork consistency protocol to detect data tampering.

More recent systems, namely VeritasDB~\cite{VeritasDB} and Concerto~\cite{Concerto}, leverage trusted hardware to speed up verification. In particular, both store Merkle tree data inside SGX enclaves. Veritas stores the roots of the trees, and Concerto uses memory verification technique to avoid contention inside the enclaves. 


\subsection{Out-of Blockchain Database}
Blockchains, which was originally designed for cryptocurrencies, is now being used as a general-purpose transactional  system. Being a distributed data processing system, a blockchain system shares some similarities with a distributed database system. However, its focus is security, whereas the database's focus is performance. The design space of both systems can be viewed along four dimensions: replication, concurrency, storage and sharding. A recent work~\cite{DBLP:journals/corr/abs-1910-01310} provides an extensive and in-depth comparison of blockchain versus database. It shows that along the four design dimensions, different choices lead to different performance.
We are seeing a trend of merging these two systems into a design that is secure, efficient, and can be readily adopted by applications such as logistic, digital banking, and digital asset management. 



One step toward realizing a hybrid blockchain-database is to support rich data queries on blockchains~\cite{BlockchainDB, BigchainDB, FlureeDB, SwarmDB}. A simple approach is to join the network as a full node and then execute the query. However, running a full node is expensive. 
vChain~\cite{vchain1} addresses this problem by embedding an aggregate and constant-size authentication data structure, constructed with multiset accumulator, in each block header.
This allows users to run a light node to query with integrity guarantee. TrustDBle~\cite{TrustDBle} proposes a secure and scalable OLTP engine that provides verifiable ACID-compliant transactions on shared data using trusted hardware.


BlockchainDB~\cite{BlockchainDB}, Veritas~\cite{Veritas},  FalconDB~\cite{FalconDB}, and LineageChain~\cite{DBLP:journals/pvldb/Ruan0DLOZ19, sigmodrh} are recent systems that use blockchain as a verifiable storage and add database features on top of it.  We now discuss these systems in more detail. 

BlockchainDB adopts a simple key-value data model, and exposes \textit{Put/Get/Verify} operations to clients. It consists of a database layer and a storage layer. The former controls the consistency level of requests so that clients can choose the balance of result staleness and performance. The storage layer serves as the unified interface to the underlying blockchains. It translates requests from the database layer into blockchain transactions and monitors the transaction status. When a client invokes \textit{Verify}, a blockchain node would contact other peers to check whether the corresponding transaction is committed in the ledger. 
A node in BlockchainDB does not hold the complete copy of the state. Instead, the states are partitioned to multiple blockchains. 

Veritas shares a similar goal and vision with BlockchainDB, but differs in three aspects. First, it targets complex data models, i.e., relational model. Second, it employs Trusted Execution Environments (TEEs) such as Intel SGX as the trustable verifiers that consume the database logs for the transaction validation. The validation results, in the form of verifiers' votes, are persisted in the blockchain. As a result, when there is a dispute, any party can resolve it by reconciling the database log with the votes on the ledger. Third, Veritas does not support partitioning, as it does not store all states on the blockchain. 

Instead of checking the ledger for log validation, FalconDB organizes database records into an authenticated data structure, such as a Merkle tree, which enables a succinct integrity proof on a database record. FalconDB employs an incentive model allowing clients to selectively challenge the transaction results from a suspicious server. If the server cannot provide proof of correctness, it will be penalized. FalconDB also supports authenticated queries on a temporal data model, so that users may access data snapshot with respect to a particular block. 

LineageChain~\cite{DBLP:journals/pvldb/Ruan0DLOZ19, sigmodrh} is a fine-grained, secure, and efficient provenance system built on top of ForkBase~\cite{DBLP:journals/pvldb/WangDLXZCCOR18, DBLP:conf/icde/LinYDCCORWXZV20} and FabricSharp\cite{fabricsharp}.  
It provides provenance information to smart contracts through simple
interfaces to enable a new class of blockchain applications whose execution logics depend on provenance information at runtime.
LineageChain captures provenance during contract execution and stores it in a Merkle tree implemented in ForkBase, and provides a novel skip list index to support efficient provenance queries.

\subsection{Ledger Database}
Amazon offers Quantum Ledger Database (QLDB)~\cite{qldb}, a cloud service that provides data immutability and verifiability. QLDB consists of blocks organized in a hash chain called journal. 
Changes to the data, including insert, update and delete, are collected into blocks and appended to the journal. A Merkle tree is built upon the entire journal. To support efficient query, the journal is materialized to user-defined tables for the latest data and history data.
QLDB aims to provide the high performance of database systems with integrity guarantees for data and historical data versions. Similarly, Oracle Blockchain Table~\cite{obt} offers append-only verifiable tables by implementing a centralized ledger model.
MongoDB~\cite{mongodb} supports verifiable change history by storing document collections in a hash chain. 

Datomic~\cite{datomic} is a distributed immutable database system designed to be ACID compliant, with datom as its database building block. It is a form of key-value store, and Datoms are collected to form an entity.  It makes use of key-value stores such as Amazon DynamoDB for managing the data, and allows users to obtain a historic snapshot of the database via its APIs and query language. Immudb~\cite{immudb} is a recently released immutable tamper-evident open source database system. Due to the demand for VDB, we foresee active development in such kind of database systems.
However, with no deletion, 
the database size will grow over time, and query processing efficiency and scalability could become major concerns. 



\section{Challenges and Opportunities}
\label{sec:designissues}
In this section, we discuss the research challenges in implementing an efficient VDB. In particular, the requirements of immutability and verifiability have implications on storage and indexing, query verification, and concurrency control mechanisms. We discuss the methodologies from recent works that provide building blocks for VDB.

\subsection{Storage and Indexing}
VDB requires an immutable and tamper-evident storage engine. In particular, the storage must support integrity proof generation, and have an efficient version management mechanism.

ForkBase~\cite{DBLP:journals/pvldb/WangDLXZCCOR18}, a storage with Git-like version control
and branch management, and Merkle-based directed acyclic graph (DAG) data structure, provides a good starting point. ForkBase supports collaborative analytics, and content-based data deduplication mechanism that significantly reduces data volume in the physical storage. Furthermore, it supports efficient version querying.

As in traditional databases, indexes are necessary for fast retrieval and location of records.
Recent Merkle tree-based indexes, namely  Merkle Patricia Trie (MPT)~\cite{Wood2014ETHEREUMAS}, Merkle Bucket Tree (MBT)~ \cite{hyperledger}, and Pattern-Oriented-Split Tree (POS-Tree)~\cite{DBLP:journals/pvldb/WangDLXZCCOR18}, support efficient queries on immutable data. 
~\cite{DBLP:conf/sigmod/YueXZCOWX20} contains a comprehensive analysis of these indices, showing that MPT, MBT, and POS-Tree are different instances of Structurally Invariant and Reusable Indexes (SIRI)~\cite{DBLP:journals/pvldb/WangDLXZCCOR18}, and that POS-tree has better overall performance. In addition to these indices which are designed for query verifiability, other indices are needed to further speed-up data retrieval.  Since versions can be modeled as temporal or historical data, indexes such as the historical R-tree\cite{DBLP:conf/sac/NascimentoS98}, and rolling index B$^x$-tree~\cite{DBLP:conf/vldb/JensenLO04} could be adapted to support the multi-dimensional and single-dimensional queries. We envision that the need for fast querying of historical data will lead to new, innovative indexes. 

\subsection{Verification}







Query verifiability in VDB means that the user who sends the query can verify the integrity of the result, that is the data and execution have not been tampered with. We discuss here different approaches to achieve verifiability. 

\textbf{Client-side verification vs Server-side verification.}
When the data is outsourced to a third party, the users themselves must verify some proofs provided by the third party. However, verification can be expensive for the users, especially when running on low-power devices. Trusted hardware, such as Intel SGX, can help mitigate this cost for user, by supporting server-side verification. In particular, the hardware performs verification securely at the servers, by running verification inside trusted execution environments, and output only succinct proofs that can be verified cheaply by the user. 
However, the secure hardware has limited resources that can lead to significant performance overhead. Furthermore, existing secure hardware are vulnerable to side-channel attacks that compromise their security. 

\textbf{Online verification vs Deferred verification.}
With respect to the timing of the verification, there are two approaches: online, and deferred verification.
In the former, the data must be committed after the verification succeeds, which is useful when recovery from malicious tampering is costly.
In the latter, verification is done over a batch of transactions, therefore achieving higher throughput than the former.

\textbf{Verification via encryption.}
One way to protect  data integrity is by using authenticated encryption.
Users can encrypt data using private key and store the ciphertexts on untrusted storage. Data tampering can be detected directly with the authentication tag. 
The limitation of this approach is that it restricts computation (or queries) on the ciphertexts.
Encryption schemes with various support for computation on ciphertexts exist, but they have trade-off in security and computation. All of these schemes have significant performance overhead. 

\textbf{Verification via authentication data structure.}
Authentication data structures, which are based on Merkle trees, provide data integrity with low cost. In this structure, the leaf nodes contain cryptographic hashes of the data blocks, while the non-leaf nodes contain the hashes of their child nodes.
The hash of the root node is called the ``digest'' of the data. 
The integrity proof consists of the hashes of the nodes from the corresponding leaf to the root of the tree. 
The new digest is recalculated recursively and equality is checked with the previously saved digest.

\subsection{Concurrency Control}
Many outsourced or cloud databases are multi-tenant. Applications running on top of a multi-tenant database may require different ACID isolation levels.
The database often has fixed the transaction isolation levels at the time of deployment, therefore applications have to implement their own levels for their needs, which increases the complexity of the system. This problem can be mitigated by using per-tenant database architecture, but this approach does not scale well.

Consider as an example an e-commerce system with customer credits.
On the one hand, the purchases of the items must occur in sequence to prevent double spending or shipping out-of-stock items.
In other words, the transaction schedule needs to be serializable, which can be implemented using optimistic concurrency control (OCC) or multi-version concurrency control (MVCC) with abortion on read-write conflicts.
On the other hand, the analysis report or status checking on the system may not require strict isolation.
Such queries are mostly processed as read-only workloads, and many of them require near real-time responses.
For example, read committed isolation will be sufficient to execute query ``getting all items with stock-level lower than 50''.
In this case, it is unnecessary to abort the query when read-write conflicts occur.

\begin{figure}[t]
    \centering
    \includegraphics[width=0.42\textwidth]{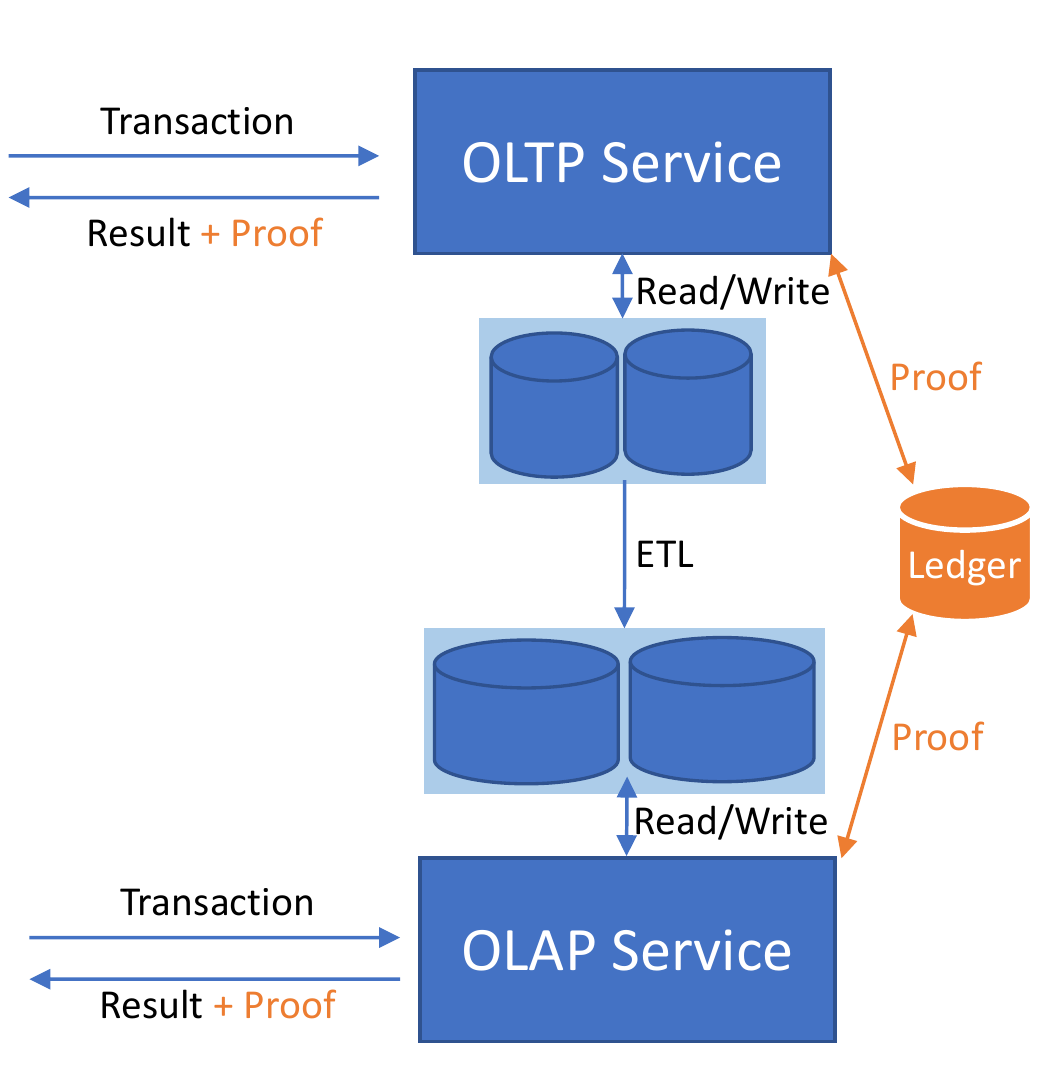}
    \caption{Non-intrusive design.}
    \label{fig:integration_solution}
\end{figure}

A common approach to achieving high performance for weak isolation is to fix the isolation to a weak level (e.g., read committed), and implement customized logic to handle stricter level in the applications.
Such design involves locks, checking pre-images of data and sometimes reversions, therefore complicating the application logic and incurring large overhead.
By providing flexible isolation levels in the underlying database,
it allows for performance optimization and lets users focus more on the application logic.

\section{Extending OLTP/OLAP to VDB}
\label{sec:extension_approach}
VDB can be implemented by adding a verifiable ledger to an existing database system. The ledger supports immutable data and verifiable queries. Here we discuss the challenges of integrating such a ledger to OLTP and OLAP systems. 

\begin{figure}[t]
    \centering
    \includegraphics[width=0.4\textwidth]{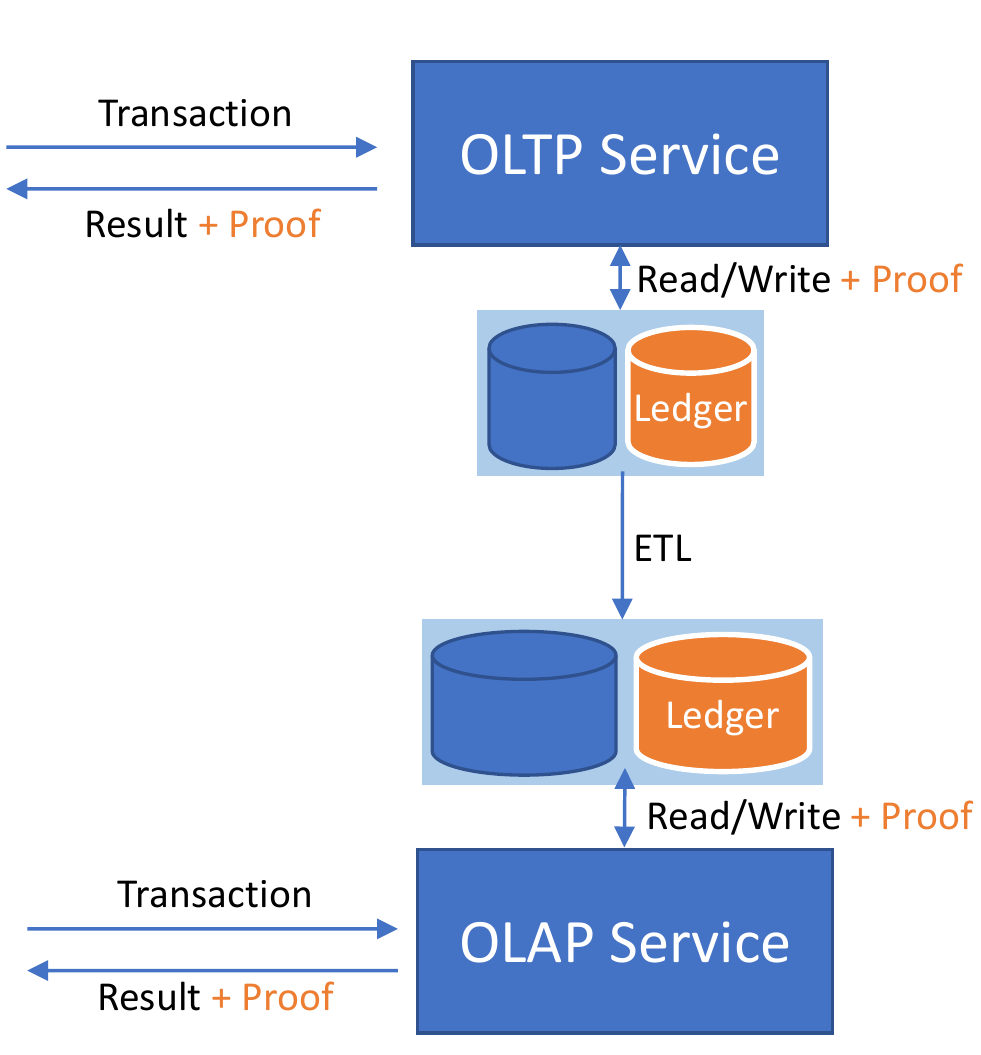}
    \caption{Intrusive design.}
    \label{fig:from_scratch_solution}
\end{figure}

There are two designs for integration, as shown in Figure~\ref{fig:integration_solution} and Figure~\ref{fig:from_scratch_solution}.
The blue arrows, rectangles and cylinders depict a typical data processing flow, where the data is collected by OLTP and analyzed by OLAP systems.


\textbf{Non-intrusive design.}
As shown in Figure~\ref{fig:integration_solution}, a ledger is attached without modifying the architecture of the original database systems. However, additional steps are added during transaction processing. The OLTP and OLAP systems generate integrity proofs from an independent ledger. On the one hand, this design minimizes disruption to existing systems, as it does not require changes to existing data.  On the other hand, it incurs considerable performance overhead, due to the interaction with the ledger. 


\textbf{Intrusive design.}
Another design, as depicted in Figure~\ref{fig:from_scratch_solution}, is to embed the ledger into an existing database system. This eliminates communication with an outside ledger, by generating the integrity proof inside the database. While reducing performance overhead compared to the other design, it incurs significant cost in data migration. In particular, data must be moved to the new system, which may be too costly for users with large amounts of data. 







\begin{figure*}[t]
\centering
\includegraphics[width=0.83\linewidth]{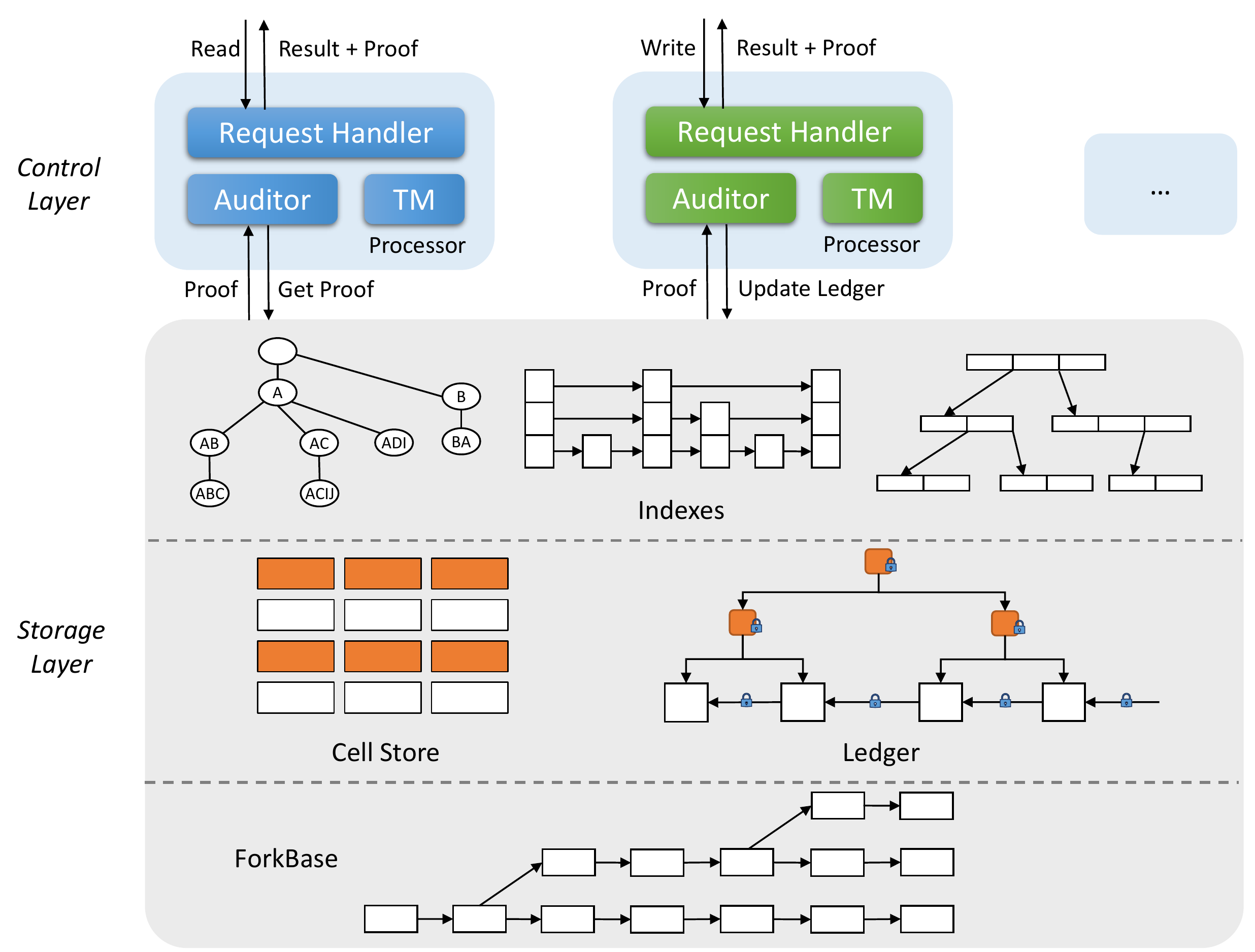}
\caption{System architecture.}
\label{fig:arch}
\end{figure*}

Another approach is to integrate the ledger with a hybrid transactional/analytical processing (HTAP) system. A HTAP system is designed to unify efficient processing of operational and analytical workloads in the same database. In the HTAP system, no data migration from OLTP system to OLAP system is necessary. Existing OLTP systems are being converted to HTAP by exploiting in-memory processing ~\cite{DBLP:journals/tkde/ZhangCOTZ15} and both columnar and row storage structures. Some recent NewSQL systems also adopt HTAP in their design and implementation.

\section{System Architecture}
\label{sec:system}
In this section, we discuss the system architecture of \systemname, a distributed database designed from scratch that supports both OLTP and OLAP workloads with verifiable ledgers. As shown in Figure~\ref{fig:arch}, the system consists of two layers: the \textit{control layer}, and the \textit{storage layer}.

The control layer consists of multiple {\em processor nodes} that accept and process requests from a global message queue\footnote{Similar to other distributed systems, the coordination, as well as the resource management, is done by a master node.}. Each node has three main components: a request handler, an auditor, and a transaction manager (TM).
The request handler accepts query requests and returns the results with the corresponding proofs.
The auditor communicates with the ledger in the storage layer to keep track of data changes. The transaction manager controls the execution of the queries in the storage.

The storage layer features a distributed storage engine, namely ForkBase. Built on top of ForkBase is a virtual cell store, as opposed to row or column store in traditional databases.
The system maps each cell to a universal key consisting of the column id, primary key, timestamp, and the hash of its value. There are multiple  index structures built into the storage layer to support verifiable query processing. 


\textbf{Ledger.} This structure consists of a sequence of hashed blocks.
Each block tracks the modification of the records, query statements, metadata and the root node of the indexes on the entire dataset.
The block and the data can be verified using the Merkle tree structure built on top of the entire ledger. Section~\ref{subsec:verification} discussed more details regarding verification and proof generation.

\textbf{Index.} \systemname uses a B$^+$-tree for query processing.
The input of the index is the requested keys, and the output is the matched data cell. This structure is efficient for both point and range queries. 

\textbf{Inverted Index.} When processing analytical queries, the system uses an inverted index to quickly locate the rows to fetch data.
Such an index uses the value recorded in each cell as index key and the universal key of the corresponding cell as value.
The structure of the inverted list varies according to the type of the data stored in the cell.
For instance, for numeric type, the system uses a skip list to better support range query, whereas for string type, it uses a radix tree to reduce space consumption.

\subsection{Query Processing}
\label{subsec:query_processing}
The processor nodes handle both read, write, and mixed workload. \systemname supports both SQL and a self-defined JSON schema.

\textbf{Write workload.}
There are four steps in handling a write workload.
(1) The request handler collects a transaction from the message queue.
(2) The auditor checks the write operations and updates the ledger. The ledger records the changes and returns a proof to the auditor.
(3) The processor traverses the B$^+$-tree index and performs the write operations to the cell store.
(4) The processor collects the results, combines them with the proof, and sends back to the user through the request handler.

\textbf{Read workload.}
The processing of read workload follows similar steps.
(1) The request handler receives a transaction from the message queue.
(2) The processor collects the results by traversing corresponding inverted indexes and retrieving the cell store.
(3) The processor visits the ledger via the auditor, getting the proofs of the results.
The proof generation is done by the ledger using the universal keys and the internal nodes of inverted index.
(4) The processor combines the results and the proofs as responses and the request handler returns them to the user.

\systemname uses a HTAP design to overcome the data movement between OLTP and OLAP systems.
Similar to the intrusive design in Figure~\ref{fig:from_scratch_solution}, it requires users to replace the underlying database systems, which might be highly tangled with their business.
However, it should be highlighted that \systemname can be used as an individual ledger by solely waking up the auditor in the processor.
Thus, the system can be applied into a non-intrusive design shown in Figure~\ref{fig:integration_solution} as a short-term transition plan of integrating \systemname into the real-world business.
Ultimately, users should use \systemname as a standalone and complete database system to cover and develop their business.

\subsection{Concurrency Control}
Concurrency control in each processor node can be implemented in the same way as in traditional database systems.
However, in our design, cells are multi-versioned. Therefore, to achieve serializability guarantee, concurrency control mechanisms based on MVCC, including MVCC with 2PL \cite{DBLP:books/aw/BernsteinHG87}, MVCC with timestamp ordering (T/O) \cite{DBLP:journals/tods/BernsteinG83}, MVCC with OCC \cite{DBLP:conf/sigmod/LimKA17}, are more suitable. 

Since each processor node processes transactions independently, it is necessary to keep the data in the indexes and the virtual storage consistent  across different nodes. The solution is to add distributed transactions to each node, and follow the two-phase commit (2PC) protocol to coordinate each transaction so that transactions committed by different nodes can be made serializable. The challenge in achieving  serializability in distributed setting is to figure out the order of transactions in the equivalently serial schedule.

One approach to achieving serializability is to rely on a global timestamp service, like Timestamp Oracle \cite{DBLP:conf/osdi/PengD10}, to allocate the timestamps upon a transaction starts and commits.
We then order transactions  based on their start timestamps.
In the prepare phase of 2PC, each transaction with read/write and write/write conflict with this order will abort.
However, there are two limitations. First, the timestamp allocation service can become the bottleneck. Second, the abort rate can be high in a write-intensive workload.
To address the first limitation, we can  adopt the hybrid logic timestamp scheme that allocates timestamps by each individual node and still has serializability guarantee \cite{kulkarni2014logical,10.1145/3318464.3386134}.
For the second limitation, it is possible to adopt the combination of OCC and MVCC by dynamically adjusting the transaction order to reduce  abort rates \cite{DBLP:conf/vldb/BoksenbaumCFP84, DBLP:journals/pvldb/MahmoudANAA14}, and verifying the transactions in batch to reduce the verification cost \cite{DBLP:journals/pvldb/DingKG18}.
These approaches 
need further investigation and evaluation.

\subsection{Proof and Verification}
\label{subsec:verification}

\systemname offers timely detection of malicious data tampering by using an authentication data structure, namely the ledger shown in Figure~\ref{fig:arch}.
Clients can use the digest of the ledger to perform verification locally.
Since changes to ledger are serializable, during the transaction processing, only the data committed before the transaction can be verified.
After the processing, clients can get the data and the proof of this transaction as described in Section~\ref{subsec:query_processing}, along with other metadata of the authentication tree structure if applicable.

To verify the correctness of the results, clients can recalculate the digest with the received proof and compare it with the previous digest saved locally.
If they match, it means the data has not been modified during the period between the verification and  when the digest is generated.
To improve verification throughput, we use a deferred scheme, which means the transactions are verified asynchronously in batch.


\section{Experimental Study}
\label{sec:exp}

In this section, we describe the prototype of \systemname and present its preliminary evaluation results.
The full-scale implementation of \systemname is in progress and a thorough performance study will be conducted in the future.

\subsection{Implementation}

First of all, we implement a baseline system to emulate a commercial product based on the features described online and testing provided by the website.
The newly inserted or modified records are collected into blocks and appended to a ledger implemented by a Merkle tree.
The ledger is used for verification purposes, shadowing the nodes of a typical B$^+$-tree for query key searching.
Furthermore, the appended blocks are materialized to indexed views for fast query processing.
To perform a read query, users can directly fetch the data with meta information using the indexed views, which can be verified against the ledger.

For the prototype of \systemname, we modify the latest version of ForkBase and forgo irrelevant functionalities such as branch management.
In particular, we implement the ledger by adopting index from Structurally Identical and Reusable Indexes (SIRI) family for both query and verification.
Each block in the ledger stores a historical index instance, naturally composing a version of the ledger, and the nodes between instances can be shared, benefiting from SIRI properties.

For comparison purpose, we also build an immutable key-value store (KVS) using ForkBase.
It is the same as \systemname in terms of indexing, except that it does not maintain a ledger or provide verifiability.
Therefore, by comparing the two systems, we can focus on the maintenance and verification cost of the ledger storage implemented in \systemname.

\subsection{Evaluation}

We evaluate the performance of the systems with read-only and write-only workloads.
The number of records, which consist of different key-value pairs, vary from 10,000 to 1,280,000.
The length of the key ranges from 5 to 12 bytes while the size of the value is 20 bytes.
The experiments are conducted on a server with Ubuntu 14.04, which is equipped with 6 cores Intel Xeon Processor E5-1650 processor (3.5GHz) and 32GB RAM.

\begin{figure*}
    \centering
    \begin{minipage}{\linewidth}
        \subfigure[Read]{
            \centering
            \includegraphics[width=0.47\linewidth]{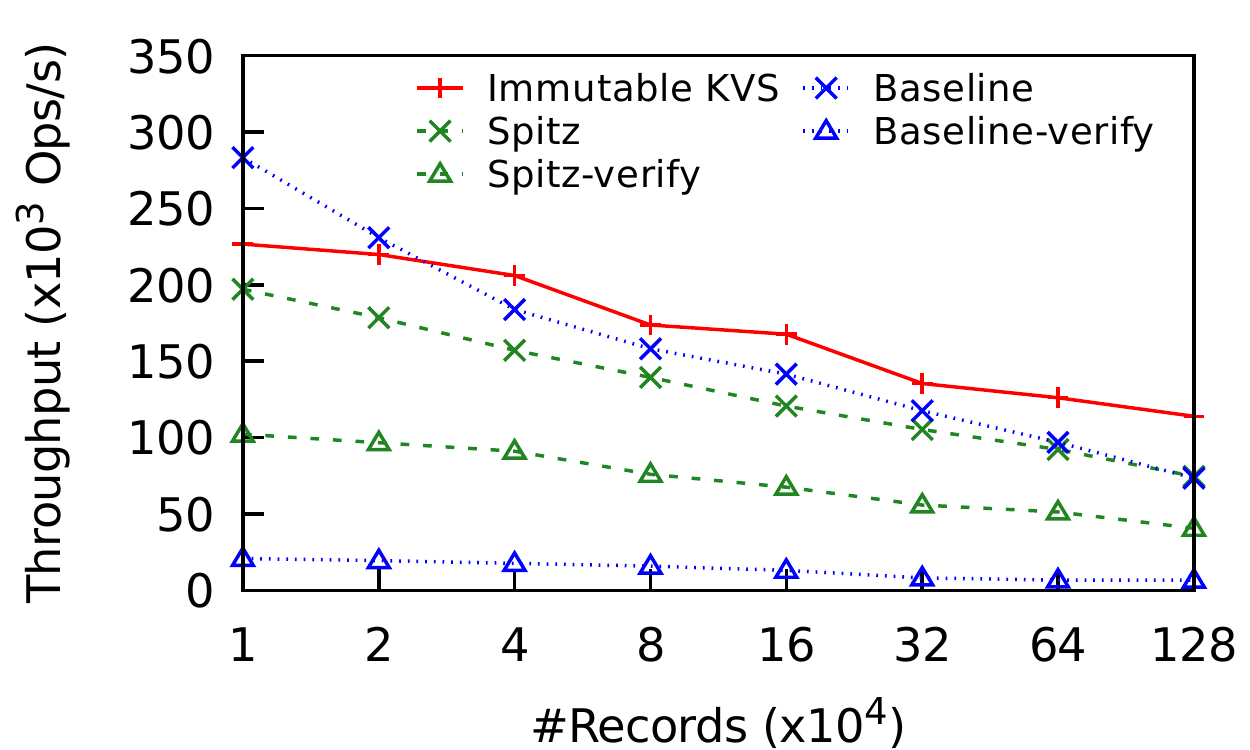}
            \label{fig:exp:basic_read}
        }
        \hspace{2em}
        \subfigure[Write]{
            \centering
            \includegraphics[width=0.47\linewidth]{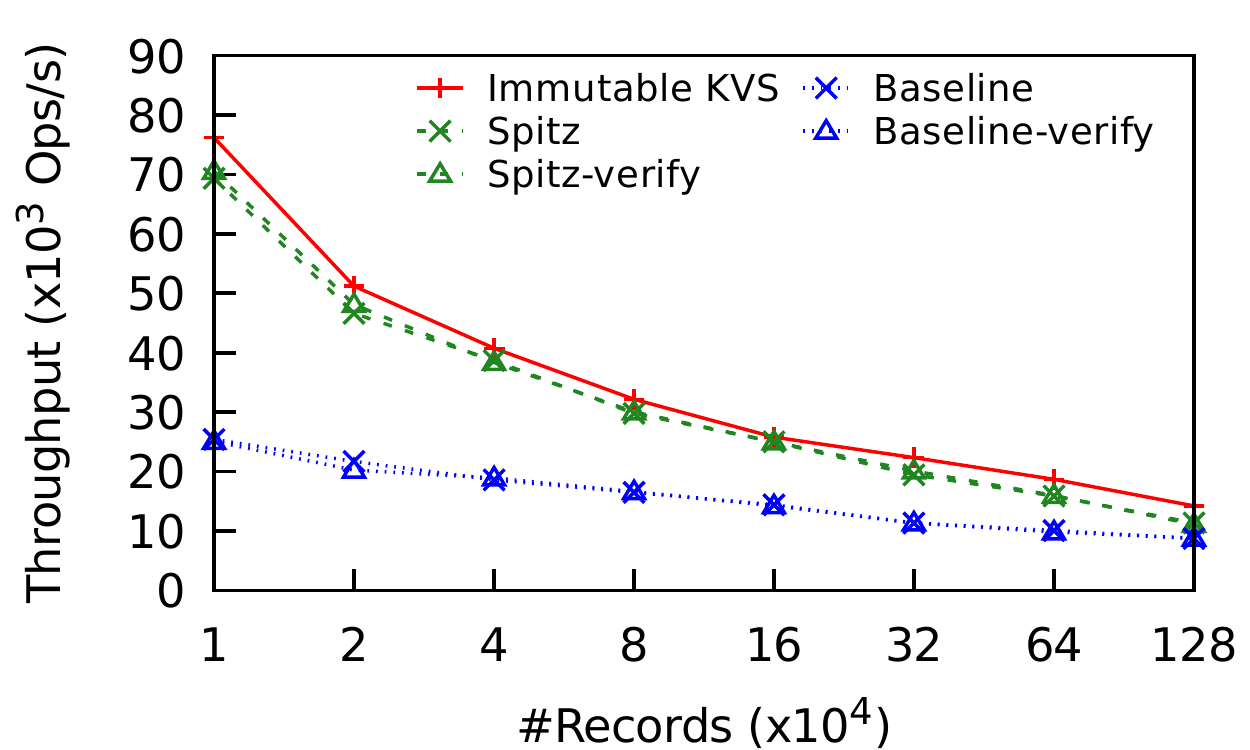}
            \label{fig:exp:basic_write}
        }
        \caption{Basic operations in single-thread setup.}
        \label{fig:exp:basic}
   \end{minipage}
\end{figure*}

\subsubsection{Basic Operations}
\label{sec:exp:basic}
We first evaluate the performance of read-only and write-only workloads in a single-thread setup.
We vary the initial database size from 10,000 to 1,280,000 records, 
and execute read-only and write-only workloads on different systems.

Figure~\ref{fig:exp:basic_read} shows the results for read-only workloads.
The immutable KVS performs the best without maintaining any verifiable data structures.
The baseline implementation and \systemname have comparable performance when the number of records becomes large, as the index traversal becomes a dominant factor in query processing.
When the verification on the integrity of the queried results is enabled, plotted as \systemname-verify and  Baseline-verify in the figure, the read performance for \systemname is approximately half of that without the verification while the baseline operations per second drops by almost two orders of magnitude.
If compared directly, 
\systemname achieves 7x operations per second than that of the baseline.
The major reason of such phenomenon is that \systemname can store the proofs of the results and the value of the target nodes in a unified index, namely the ledger implemented via SIRI.
To compare, the baseline needs to visit the B$^+$-index first, and uses the resultant nodes to get the proof from the ledger.
Figure~\ref{fig:exp:basic_write} shows the results for write-only workloads.
Similarly, thanks to the unified index structure, \systemname has operations per second comparable to the immutable KVS with and without verification while the performance of the baseline system is much worse because of maintaining multiple indexed views.

\subsubsection{Range Query}

\begin{figure}
    \centering
    \includegraphics[width=\linewidth]{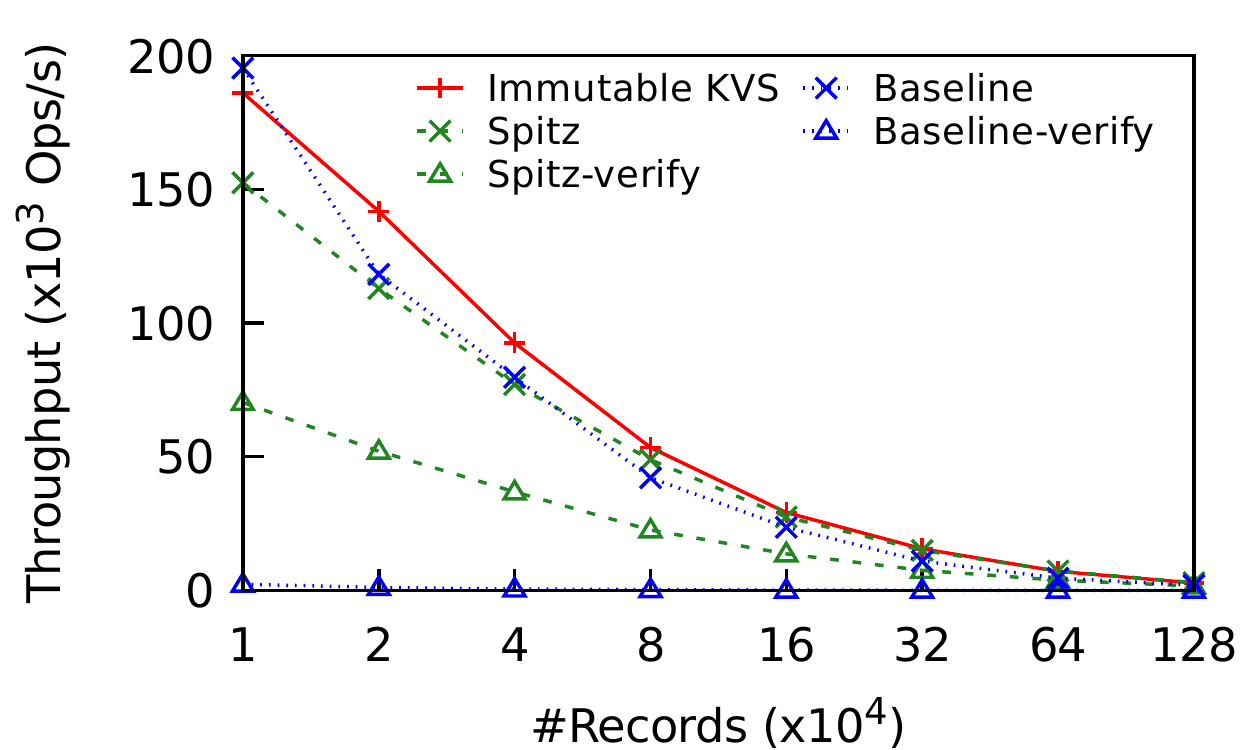}
    \caption{Range query performance.}
    \label{fig:exp:range}
\end{figure}

In this section, we evaluate the performance of analytical workloads with range queries.
Such workloads are commonly submitted by data scientists to retrieve a group of records for analysis or further aggregation.
We initialize the database with 10,000 to 1,280,000 records for different runs.
The selection conditions of the range query are set on the primary key and the selectivity of the query is fixed at 0.1\%.

Figure~\ref{fig:exp:range} depicts the operations per second in all systems.
As can be seen, the performance of the range query is worse than the performance of point query shown in Figure~\ref{fig:exp:basic_read} by 25\% to 90\% for all cases.
This is due to the additional nodes needed to be traversed and scanned when the query is processed.
Meanwhile, the operations per second drops fast when the total number of records increases because the systems fetch increasing number of records with fixed selectivity for all cases.

The baseline stores the ledger and the index of the data separately, and hence, the retrieval of the proofs cannot benefit from the optimizations used in range query processing.
That is, the retrieval on the proofs of resultant records, instead of being fetched in a batch by scanning keys with the given interval, must be processed by searching the digest in the ledger individually.
In contrast, for \systemname, thanks to the use of the unified index structure described in Section~\ref{sec:exp:basic}, proof retrieval can leverage the traversal on the index of the data -- the proofs of the resultant records are returned simultaneously when the resultant records are scanned and selected.
Consequently, for queries with verification of data integrity enabled, \systemname outperforms the baseline by up to two orders of magnitude, a gap much larger 
than the results shown in Figure~\ref{fig:exp:basic_read}.


\subsubsection{Non-intrusive Design vs \systemname}

\begin{figure*}
    \centering
    \begin{minipage}{\linewidth}
        \subfigure[Read]{
            \centering
            \includegraphics[width=0.47\linewidth]{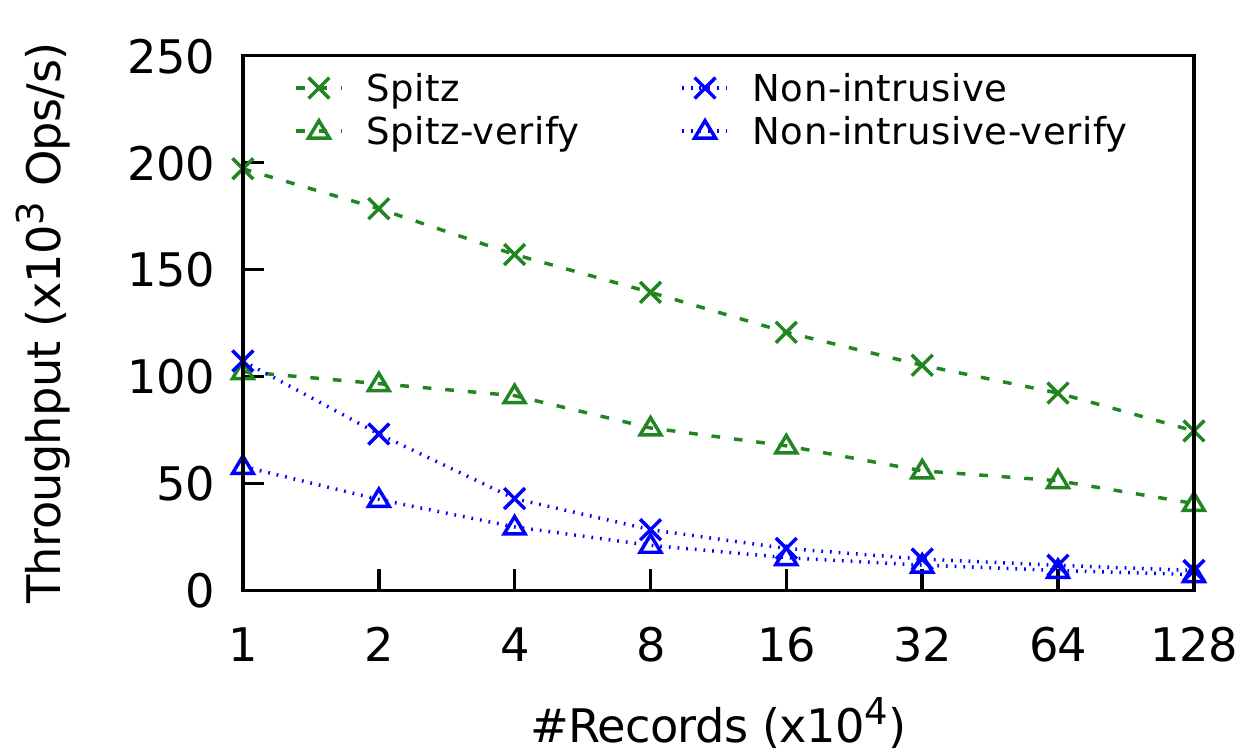}
            \label{fig:exp:non_intrusive_read}
        }
        \hspace{2em}
        \subfigure[Write]{
            \centering
            \includegraphics[width=0.47\linewidth]{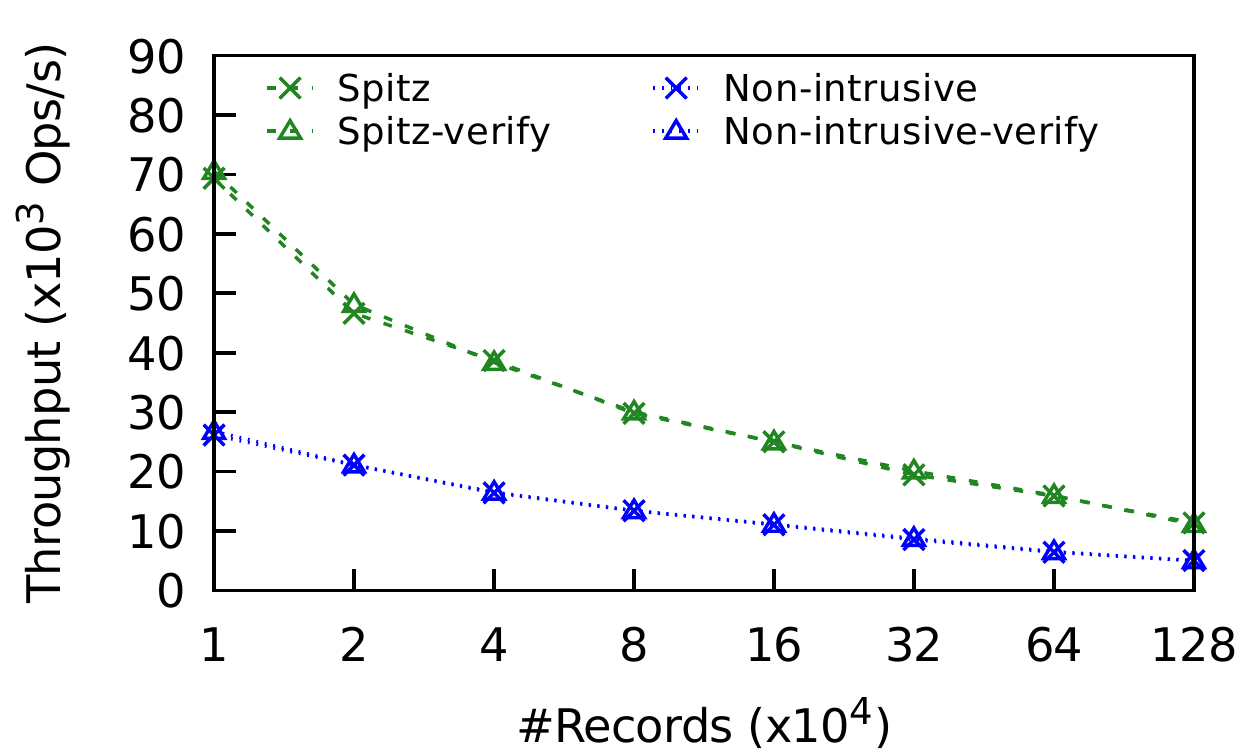}
            \label{fig:exp:non_intrusive_write}
        }
        \caption{Non-intrusive design vs. \systemname.}
        \label{fig:exp:non_intrusive}
   \end{minipage}
\end{figure*}

In this section, we evaluate the performance of a non-intrusive design of VDB and compare it with \systemname.
We set up an immutable key-value store using ForkBase as the underlying system, which interacts with the ledger shown in Figure~\ref{fig:integration_solution}.\footnote{ForkBase can be treated as a HTAP system here therefore we do not initialize separate OLAP and OLTP system as shown in the figure.}
To support data verification, we deploy \systemname on the same server as the Ledger database in the figure.
In the case of read workloads, the client obtains the queried results from the underlying database and the proofs from the ledger as responses, while in the case of write workloads, the submitted data are committed in both the underlying and ledger database atomically.
To verify the results, the client uses the proof from the Ledger database, calculates the digest of the returned results, and compares it with the previous digests as described in Section~\ref{subsec:verification}.

The experiment is conducted with read-only and write-only workloads and the results are shown in Figure~\ref{fig:exp:non_intrusive}.
As can be seen, the non-intrusive design incurs significant overhead by maintaining two systems, i.e., the underlying database and the Ledger database.
Specifically, for read-only workloads, the performance of \systemname is 6x higher than the non-intrusive design when the verification of data integrity is enabled.
The huge performance gain comes from a simpler process flow:
the request can be processed within a single system in \systemname, while in the non-intrusive design, it must be sent to the underlying database first to obtain the results, and passed to the Ledger database to retrieve the proofs.
Obviously, the interactions between the Ledger database and the underlying database inevitably introduce additional cost on network communication, query planning, etc.
For write workloads, \systemname produces 3x higher number of operations per second
than the non-intrusive design.

In summary, the results show that the prototype of \systemname achieves better performance than the non-intrusive design. 
Without doubt, its performance could further be improved with indexes and optimization strategies specifically designed for VDB.




\section{AI and VDB - A Synergy}\label{ai}

We have discussed the design and implementation of a VDB that can support new database applications.  
This new trend in database coincides with the rapid development of artificial intelligence (AI). AI has been proven successful in a wide range of applications, could automate many tasks, and it is often better at modeling complex situations than humans.

To  meet  the  demand for complex analytics, database systems have enabled the addition of machine (or deep) learning libraries to construct end-to-end analytics pipelines.
With the evolution of dataset due to versioning, machine learning models used in the analytics pipeline may exhibit concept drift behavior, which causes the model to become less accurate over time.
Therefore, iterative analytics component updates may become necessary, and the relationship between component and data versions need to be maintained for verification on the analytical results.

From system performance perspective, deep learning can be used to enhance database performance and usability, and vice versa,  deep learning can benefit from the efficient data management and performance provided by databases (Apache SINGA~\cite{DBLP:conf/mm/OoiTWWCCGLTWXZZ15} for example).
Earlier works~\cite{DBLP:journals/sigmod/0059Z0JOT16} have discussed the symbiotic relationship between databases and machine learning. 
In the following, we shall continue this discussion by making a case for the merging of VDB and AI.

\subsection{AI for VDB}
AI can make VDB more intelligent. Currently, VDB design is based on empirical methodologies, which might not have a good performance. 
As pointed out in~\cite{DBLP:journals/sigmod/0059Z0JOT16}, AI may help to improve VDB’s performance in several aspects. 
\begin{itemize}
    \item Learning-based data structure. A number of solutions investigate how to enhance existing indexes or design new indexes for better storage and query efficiency, e.g., learned B$^+$tree~\cite{DBLP:conf/sigmod/KraskaBCDP18}, secondary index~\cite{DBLP:conf/sigmod/WuYTSB19}.
    \item Learning-based transaction management. AI could be used to predict the future transaction workload~\cite{DBLP:conf/sigmod/MaAHMPG18} and schedule the transactions such that the throughput is maximized with an acceptable abort rate~\cite{DBLP:journals/corr/abs-1903-02990}.
    \item Learning-based performance tuning. To avoid manual tuning of the memory allocation or I/O control, recent works~\cite{DBLP:conf/sigmod/ZhangLZLXCXWCLR19} apply reinforcement learning to automatically tune database configurations according to workload changes.
    \item Learning-based query optimization. To optimize the queries, existing works~\cite{DBLP:journals/corr/abs-1905-06425, DBLP:journals/pvldb/YangLKWDCAHKS19} deploy deep neural networks such as convolutional neural networks (CNNs), recurrent neural networks (RNNs) and their variants to estimate the cardinality and cost. 
\end{itemize}

Though many learning-based techniques for general DB have been proposed, more effort is needed to adapt them for VDB, since VDB has different data structure, storage and transaction management requirements.

\subsection{VDB for AI}
VDB can make AI (-based analytics) more reliable. 
One key feature of VDB is that it maintains historical data. 
Hence, a natural question might be: can VDB support analytical queries? For example, a client (e.g., a hospital) has already outsourced its database to a cloud hosting company for processing online transactions (e.g., medical records).  It may then also want the VDB to process some analytical queries (even some machine learning model) for specific evaluations, such that it does not need to download the data and execute locally. 
However, the analytical query support in VDB is limited (e.g.~\cite{QLDB-Querying}). 
More importantly, the analytics result should be verifiable, ensuring that it is computed from correct data; otherwise, it may result in a wrong decision, and lead to huge loss (e.g., could be people’s health or even life in medical domain).

There are several works~\cite{DBLP:conf/ccs/ZhangKP15, ZhangGKPP17} that support verifying arbitrary SQL queries over outsourced databases, but with low  performance. 
Recent works~\cite{vchain1, DBLP:conf/icde/Zhang0XTC19} on verifiable specific queries over blockchain (can be viewed as a special kind of VDB) are relatively efficient, but they only support range queries. It is necessary
to investigate how to efficiently compute complex analytical queries on VDB.

\begin{figure}
    \centering
    \includegraphics[width=0.84\linewidth]{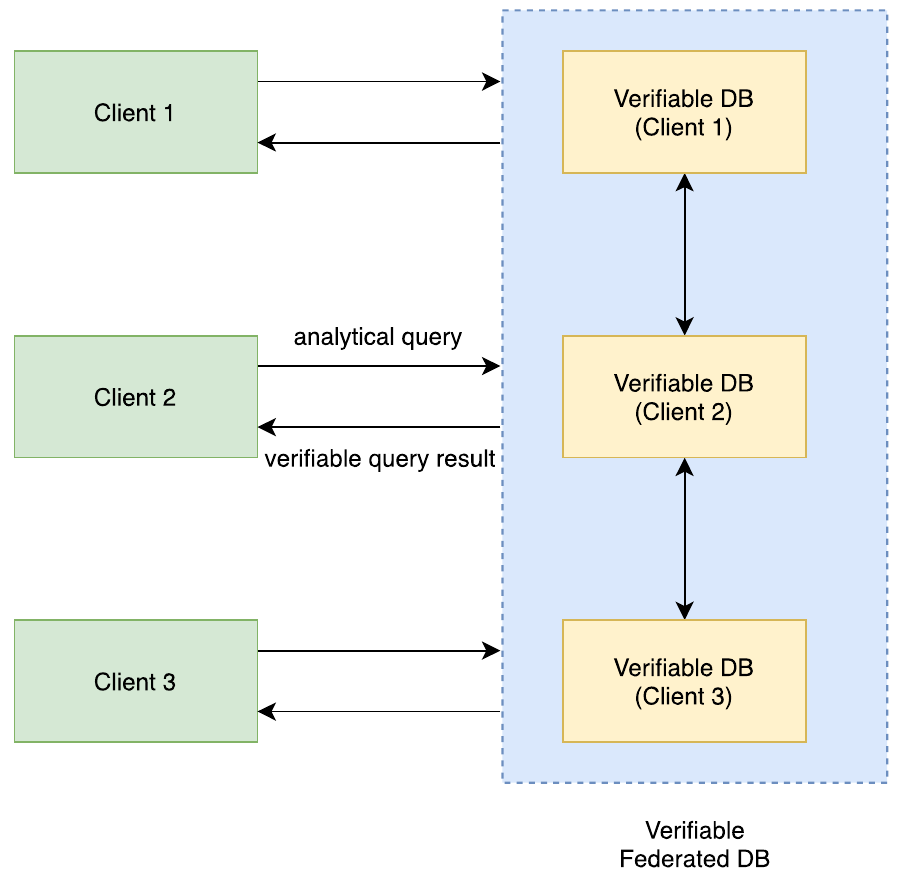}
    \caption{Verifiable federated analytical query processing.}
    \label{fig:federated_analytics}
\end{figure}

Finally, it is possible to consolidate multiple clients’ VDB to provide federated analytics (as illustrated in Figure~\ref{fig:federated_analytics}).  For example, a few hospitals want to have a more precise and comprehensive analysis of a disease. The integrity of the data and queries are important in these use cases. At the same time, each client should not be able to break the confidentiality of the other clients’ data.

In summary, AI and VDB could benefit from each other: AI could improve the performance of VDB, and VDB could ensure the trustworthiness of the analytical results (or AI models).

\section{Conclusions}
\label{sec:conclude}
With recent digital optimization and transformation, more and more businesses are transacting directly with each other.  
The current pandemic further speeds up the transformation and adoption of online business processes. 
This trend introduces a new important requirement to database systems:
the integrity of the data, the history, and the execution must be protected. This gives rise to a new class of database systems that support the verification of the transactional integrity.  

In this paper, we discuss the requirements and challenges of verifiable databases. We present approaches to extend existing systems to support verification, and an initial design and prototype of our ongoing development of a new  VDB system called \systemname.  We conduct an experimental study and show that \systemname is able to provide a better performance than a baseline system.  As future works, we will continue to implement the system and present a comprehensive system design and a thorough performance study. We will also study and possibly introduce new indexes, concurrency control mechanisms and query processing strategies for VDB.


\balance

\section*{Acknowledgments}  

Meihui Zhang would like to thank the VLDB Endowment Awards Selection Committee for the 2020 VLDB Early Career Research Contribution Award, and the nominators for the nomination.
After checking with the Chair of the Awards Selection Committee on the invited paper requirements, Meihui decided to report an ongoing system development, which is being built upon system components and works developed by her and collaborators.   For the works that led to the award, Meihui would like to thank her mentors, colleagues, collaborators, research assistants and students for their contributions. Meihui would also like to thank her ex-dean, Prof. Heyan Huang and current dean Prof. Guoren Wang, for their support and guidance.

For this paper, Meihui, Zhongle, Cong and Ziyue would like to thank Anh Dinh, Qian Lin, Wei Lu, Beng Chin Ooi, Pingcheng Ruan, and Yuncheng Wu for their discussions, contributions and proof reading.
The research of Cong Yue and Zhongle Xie are supported by Singapore Ministry of Education Academic Research Fund Tier 3 under MOEs official grant number MOE2017-T3-1-007.

\bibliographystyle{abbrv}
\bibliography{vldb} 


\end{document}